 \documentclass[smallabstract,smallcaptions]{dccpaper}

\usepackage{epsfig}
\usepackage{amsmath}
\usepackage{amssymb}
\usepackage{color}
\usepackage{url}
\usepackage{graphicx}
\usepackage{subfig}

\newlength{\figurewidth}
\newlength{\smallfigurewidth}

\newcommand{\etal}{\textit{et al.}}
\begin{document}

\title
{\large
\textbf{Transformer-based Image Compression}
}

% \author{%
% Author One$^{\ast}$, Author Two$^{\dag}$, and Author Three$^{\ast}$\\[0.5em]
% {\small\begin{minipage}{\linewidth}\begin{center}
% \begin{tabular}{ccc}
% $^{\ast}$Institution One & \hspace*{0.5in} & $^{\dag}$Institution Two \\
% Street Address One && Street Address Two \\
% City, State, ZIP, Country && City, State, ZIP, Country\\
% \url{email@address} && \url{email@address}
% \end{tabular}
% \end{center}\end{minipage}}
% }

\author{%
Ming Lu$^\dag$, Peiyao Guo$^\dag$, Huiqing Shi$^\ddag$, Chuntong Cao$^\ddag$, and Zhan Ma$^\dag$ \\[0.1em]
{\small\begin{minipage}{\linewidth}\begin{center}
\begin{tabular}{cc}
\multicolumn{2}{c}{$^\dag$Nanjing University, $^\ddag$Jiangsu Longyuan Zhenhua Marine Engineering Co.} \\
%\url{luming@smail.nju.edu.cn} & \url{mazhan@nju.edu.cn}
\end{tabular}
\end{center}\end{minipage}}
\\[0.2em]
}

\maketitle
\thispagestyle{empty}

\begin{abstract}
A Transformer-based Image Compression (TIC) approach is developed which reuses the canonical variational autoencoder (VAE) architecture with paired main and hyper encoder-decoders. Both main and hyper encoders are comprised of a sequence of neural transformation units (NTUs) to analyse and aggregate important information for more compact representation of input image, while the decoders mirror the encoder-side operations to generate pixel-domain image reconstruction from the compressed bitstream. Each NTU is consist of a Swin Transformer Block (STB) and a convolutional layer (Conv) to best embed both long-range and short-range information; In the meantime, a casual attention module (CAM) is devised for adaptive context modeling of latent features to utilize both hyper and autoregressive priors.
%By embedding the resolution re-scaling in convolutionayers, STBs can best capture the nonlocal attention for better exploring the spatial correlation. 
%Deep learning based image compression algorithms have achieved impressive performance leading against traditional methods. The convolutional neural networks are generally utilized as the backbone of both the encoder and decoder networks due to its powerful modeling capabilities. However, the locality of convolution often limits the nonlocal information aggregation, and the weight sharing of kernels hinders the content adaption of the model. To address that, inspired by recent emerging vision transformer networks, we propose a transformer-based image compression method. By establishing  correlation coefficient of tokenized input images and causal attention of context, 
The TIC rivals with state-of-the-art approaches including deep convolutional neural networks (CNNs) based learnt image coding (LIC) methods and handcrafted rules-based intra profile of recently-approved Versatile Video Coding (VVC) standard, and requires much less model parameters, e.g., up to 45\% reduction to leading-performance LIC.

%To the best of our knowledge, our model is the first transformer-based image compression method.

\end{abstract}

\section{Introduction}
High-efficiency image compression plays a vital role for effective Internet service, such as the online advertisements, professional photography sharing, etc. Though traditional rules-based image coding standards, e.g., JPEG~\cite{wallace1992jpeg}, JPEG2000~\cite{rabbani2002jpeg2000}, Video Coding Intra Profile, etc, are widely deployed for decades, a better image coding approach, i.e., less bandwidth consumption but better reconstruction quality, is a constant desire for better service. 

Recent years, a variety of deep learning based image compression approaches~\cite{balle2016end,balle2018variational,minnen2018joint,cheng2020learned,chen2021end} have emerged with noticeable coding efficiency improvement, even offering better performance than the VVC Intra~\cite{bross2021overview} quantitatively and qualitatively~\cite{cheng2020learned,chen2021end}. 
% In addition, the same LIC methods could be extended to quickly support different image sources (e.g., RGB, YUV, Bayer RAW, etc)~\cite{Zhihao_RAW},  directly infer content semantics in compressed feature domain without pixel decoding (e.g., object classification, detection, etc)~\cite{codedvision}, and flexibly execute task-oriented optimization (e.g., object-based coding, perceptual quality fine-tuning, etc)~\cite{xia2020object, liu2018deep, Wang_2021_CVPR}. 
% All of these attractive functionalities
This attractive potentiality encourages the pursuit of next-generation image coding techniques from both academia researchers and industrial leaders. Thus, international standardization groups, such as the well-known ISO/IEC JPEG, have officially called for the technical evaluation and standardization of deep learning based image compression.

\subsection{Background and Motivation}

As seen, existing LICs generally use CNNs in VAE framework where a number of convolutional layers (with resolution scaling) are stacked to exploit spatial correlation in local neighborhood to derive compact latent features  at bottleneck for entropy coding. Applying CNNs for discriminative feature extraction was originally inspired by the discoveries of Hubel and Wiesel about the cats’ visual cortex in 1960s~\cite{hubel1962receptive}.

However, the CNNs do have limitations~\cite{liang2021swinir}. First, convolutional filter  can only characterize short-range spatial correlation within the receptive field; Then, offline trained CNN model uses fixed parameters, making it generally incapable of dealing with images having very different content distribution; Third but not the last, CNN computation is computational intensive due to element-by-element processing.

To overcome the drawbacks of native CNNs, a number of principled rules are then developed. For example a variety of attention mechanisms, e.g., nonlocal, spatial- and channel-wise methods, are manually integrated with convolutions to capture long-range correlations for more compact latent feature with better compression performance~\cite{chen2021end, cheng2020learned}. However, these handcrafted attention methods often require excessive model parameters which makes them impractical for real-life applications. To deal with images with different content distributions,  multi-model optimization helps the LIC encoder to select an optimal one with better rate-distortion measurements~\cite{lu2020end}, which apparently requires us to cache multiple models.

Recalling that the human visual system (HVS) usually scans the natural environment during a saccade to fixate on different regions for saliency extraction and scene understanding, it helps us to effectively capture long-range correlations for better information embedding. 
Such biological visual fixation with attentive window adaptation is well simulated by the shifted window method  suggested in award-winning Swin Transformer\cite{liu2021swin}. And, having the successful deployment of convolutions  in existing LICs to embed local neighborhood information, this work suggests to combine the  convolutional layer (for short-range information embedding) and Swin Transformer-based attention block (for long-range information aggregation), by which we wish to produce more compact latent features for better coding efficiency.

%Aforementioned handcrafted attention mechanisms have already shown the potential along this direction. 
%On the other hand, recently-emerged vision Transformers have shown the superior performance in many tasks.
%Thus, this work puts a step forward by combining the  convolutional layer (for short-range information embedding) and attention block (for long-range information aggregation). Instead of applying handcrafted attentions as in~\cite{chen2021end, cheng2020learned}, we suggest to use the award-winning Swin Transformer to automatically and intelligently examine and aggregate long-range and necessary  information. In this way, we wish to produce more compact latent features for better coding efficiency.

\subsection{Our Approach and Contribution}

This paper proposes the Transformer-based Image Compression (TIC) method. The TIC applies the same VAE architecture as used in existing LICs. For better information embedding, we suggest the neural transformation unit (NTU) as the basic module which is comprised of a Swin Transformer block (STB) and a convolutional layer (Conv).  Similarly, we use the resolution scaling in convolutional layer (with predefined strides) to characterize and embed spatial information. To efficiently encode the latent feature at the bottleneck, both hyper priors and autoregressive neighbors are utilized as in~\cite{minnen2018joint,cheng2020learned,chen2021end} through a causal attention module (CAM) that combines the causal self-attention  and the multi-layer perceptron (MLP) to exploit closely-related priors for context modeling.

%Its basic processing unit includes a convolutional layer and a STB unit. 

%according to the paradigm of transformer networks. The input image is tokenized and scaled using convolutional layers, followed by the Swin Transformer blocks served as the backbone to build up the encoder and decoder networks for global information aggregation. A causal attention module consist of a causal self-attention mechanism and several multi-layer perceptrons (MLP) is adopted for adaptive context entropy coding. 

Experimental results show that the proposed TIC shows competitive compression efficiency , and only requires about a half of model parameters to  the state-of-the-art LIC method - Cheng {\it et al.}~\cite{cheng2020learned}. We then extend the TIC by placing more STBs in layers closer to  the bottleneck at main coder,  denoted as the TIC+, by which we can even outperform the VVC Intra with a slight increase of model parameters.

%using model with more parameters namely TIC+ proves the state-of-the-art performance, which is even better than the intra profile of VVC with xxx BD-rate gains.    

%\subsection{Contribution}
%To summarize, our contributions are listed as below. 1) We propose a transformer based image compression method to improve the compression efficiency. 2) Swin Transformer is utilized to build up an auto-encoder network for global information aggregation. 3) A causal attention module is used for autoregressive prediction in context entropy coding.

\section{Related Work}

{\bf Learnt Image Compression.} The core problem of image compression is about the compact representation of pixel blocks, for which over decades, the transforms, such as the Discrete Cosine Transform, Wavelet, etc, and context adaptive entropy codes have been extensively studied to fulfill the purpose.

Back to 2016, Ballé et al.~\cite{balle2016end} showed that stacked convolutional layers can replace the traditional  transforms to form an end-to-end trainable image compression method with better efficiency than the JPEG. Then, by adopting the hyper prior~\cite{balle2018variational} and autoregressive neighbors~\cite{minnen2018joint} for entropy context modeling, the image compression efficiency was further improved and was competitive to the Intra Profile of the High-Efficiency Video Coding (HEVC). Cheng \etal~\cite{cheng2020learned} later introduced a Gaussian mixture model for better approximating the distribution of latent features, with  which the comparable performance to the VVC Intra was reported. In addition to these methods mainly utilizing the CNNs to analyze and aggregate information locally, our early exploration on neural image coding (NIC) in~\cite{chen2021end} applied a nonlocal attention to the intermediate features generated by each convolutional layer to select attentive features for more compact representation. 
%\gpy{ which is selected by the IEEE 1857.11 committee as the baseline model for standardizing the next-generation image compression technology because of its superior coding efficiency.} 
However, the nonlocal computation is expensive since it typically requires a large amount of space to host a correlation matrix with size of $HW\times HW$ for a naive implementation. Note that $H$ and $W$ are the height and width for input feature map.

As seen, handcrafted attention mechanisms have already promised the  potentials for improving the coding efficiency~\cite{cheng2020learned,chen2021end}, in which they are mostly utilized  to  guide the encoding of importance area. 
Whereas, this work suggests to combine the convolution and attention to jointly consider the short-range and long-range correlation effectively.

%{\textcolor{red}{having NLN in hyper is bad for poor performance. How about TIC for small size image?}}
%capture long-range correlation between features for better information embedding. However, they still relied on the CNNs to generate intermediate features and only used non-local attentions to

%used CNN as the embedding layers for feature extraction and the attention coefficient is calculated using the entire input at the smallest size, which is different from ours.    

{\bf Vision Transformer.} Since the AlexNet, the CNNs have improved the performance of numerous vision tasks remarkably, by relying on the powerful capacity of stacked convolutions in characterizing the input data. Even though, the locality of the convolutional computation limits its performance by only aggregating the information within the limited receptive field. Recently, emerging vision transformer networks~\cite{carion2020end,dosovitskiy2020image,zhou2021deepvit} have demonstrated very encouraging improvement to those methods only using CNNs  for various high-level vision tasks. One potential reason is that the Transformer structure can well capture the long-range correlation for better information embedding, by  dividing the input images into tokenized patches and treating them as sequential words as in natural language processing  for computation~\cite{vaswani2017attention}. 

Later, a number of improvements have been made, including the position encoding, token pooling, feed forward networks and layer normalization, to further the performance. Particularly, the emergence of the award-winning Swin Transformer~\cite{liu2021swin} has revealed that the Transformer can be also extended to low-level vision tasks~\cite{wang2021uformer,liang2021swinir} for better reconstructed quality.  The Swin Transformer applies the window-based attention and relative position encoding,  making it capable of processing arbitrary-size  high resolution images. Applying the computation on non-overlapped windows can significantly reduce the computational complexity  when compared with the normal convolutions that traverses all elements in a frame. In addition, the shifted window scheme performs the nonlocal processing at the patch-sized level to exploit the long-range dependency for better information aggregation.

%Given that handcrafted attention already promises its potentials in image/video coding~\cite{chen2021end}, 
This work attempts to migrate the Swin Transformer into the image compression pipeline for better performance.

\section{Method}

\begin{figure}[t]
\begin{center}
\subfloat[][]{
\label{fig:basic}
\includegraphics[width=0.28\linewidth]{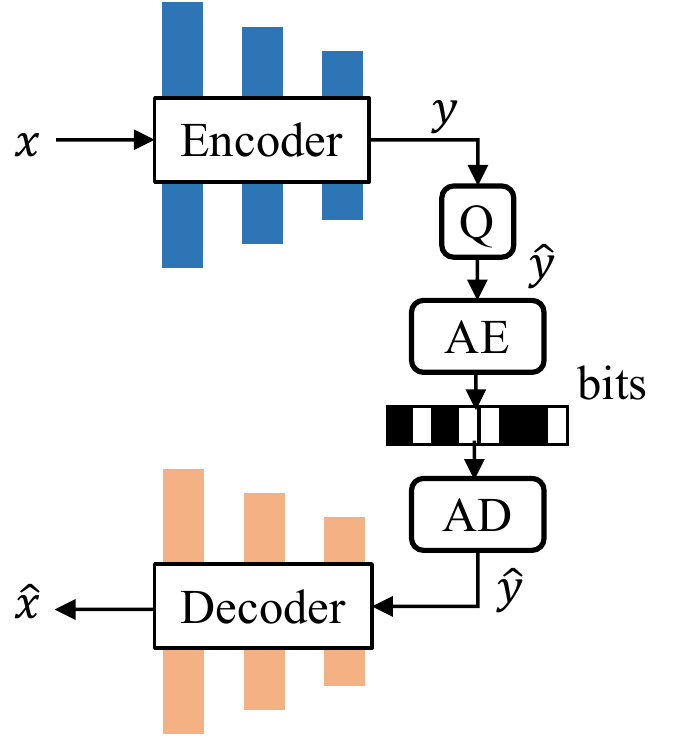}
}
\subfloat[][]{
\label{fig:lic}
\includegraphics[width=0.33\linewidth]{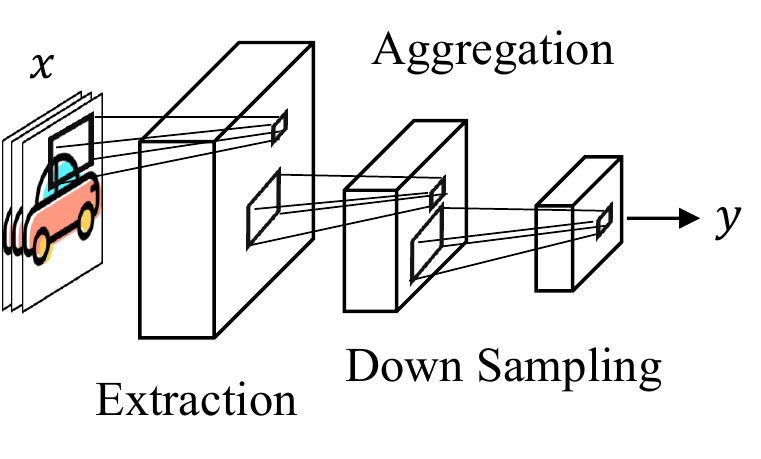}
}
\subfloat[][]{
\label{fig:tic}
\includegraphics[width=0.33\linewidth]{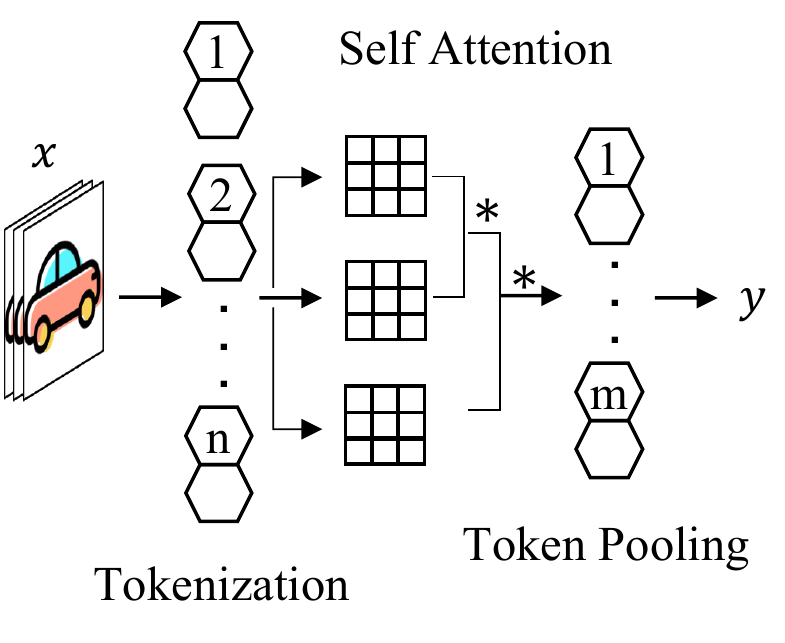}
}
\end{center}
\caption{\label{fig:pipeline}%
(a) The common paradigm of end-to-end learning-based image compression. Q, AE and AD represent the quantization, arithmetic encoding and decoding respectively. (b) The encoding flow of CNN-based image compression. (c) The encoding folow of Transformer-based image compression.}
\end{figure}

\subsection{Local Convolution and NonLocal Transformer for Image Compression}

Figure~\ref{fig:basic} illustrates the common paradigm of end-to-end learning-based image compression methods.
% The encoder is used to transform the input $x$ into the latent features $y$, and $y$ is then quantized to obtain the discrete latent features $\hat{y}$, which is then losslessly compressed into bitstreams using entropy coding like arithmetic coding.
{The encoder transforms the input $x$ into the latent features $y$, and then quantize $y$ to discrete symbols $\hat{y}$ that are entropy coded into bitstreams using predefined distribution models; The decoding part mirrors the encoding steps by parsing the compressed bitstream to reconstruct pixel blocks to form decoded  $\hat{x}$. }
% The inverse operations are deployed for decoding to transform the quantized features into the decoded image $\hat{x}$. 
The optimization objective is to minimize the rate-distortion cost through an end-to-end learning means:
\begin{equation}
    L = R(\hat{y}) + \lambda D(x, \hat{x}),
\label{eq:rd}
\end{equation}
where $R$ is compressed bit rate of $\hat{y}$ and the distortion $D$ measures the mean square error (MSE) between the ground truth $x$ and restored output $\hat{x}$.
% & $D=MSE(x, \hat{x})$ in this paper. 
We adapt $\lambda$ for rate-distortion trade-off at various bit rates.

Figure~\ref{fig:lic} details the encoding flow of existing LICs that mainly use CNNs for image coding. It uses stacked convolutional layers to analyze and aggregate features for compact representation of the input at the bottleneck. Often times, convolution is coupled with predefined strides for resolution scaling and spatial neighborhood information embedding. As seen, the receptive field can be enlarged by the resolution scaling to exploit spatial correlation from more local neighbors. 
At the bottleneck, quantized latent features are then entropy-coded through  context models conditioned on autoregressive neighbors and hyper priors.  To jointly utilizing the autoregressive neighbors and hyper priors for better context modeling, masked CNN~\cite{salimans2017pixelcnn++} is typically used to fuse information locally. 

% The convolutional layer is utilized as the extractor of the network to extract the features of input. By setting the stride size at 2 or more, features can also be down sampled using the convolutional layers. With such stacked convolutional layers, the information of the features is aggregated for further entropy coding. 
%\gpy{(Here needs to describe the difference between transformer-based and convolution-based. And in the final, which is somehow similar to the vision transform network. The paragraph below should be revised for better understanding.)}
Additionally, Figure~\ref{fig:tic} shows the encoding flow by using the Transformer. Input image is first tokenized using a convolutional layer to produce fixed-size tokens (i.e., feature patches after sequential projection); Then self attention layer is applied to derive spatial coefficients for upcoming token pooling by which we can intentionally remove less important tokens.

This work attempts to leverage the advantages of both local convolution and Transformer-based nonlocal attention for better information embedding.

%The transformer-based encoder is somehow similar to the vision transformer network as shown in . Features are often divided into patches with fixed size and then projected into sequential tokens. are utilized for spatial coefficient calculation. The token pooling operation is also deployed to reduce the unimportant tokens.  

\subsection{TIC: Exploring the Combination of Convolution \& Transformer}

\begin{figure}[t]
\begin{center}
\begin{tabular}{c}
\epsfig{width=5in,file=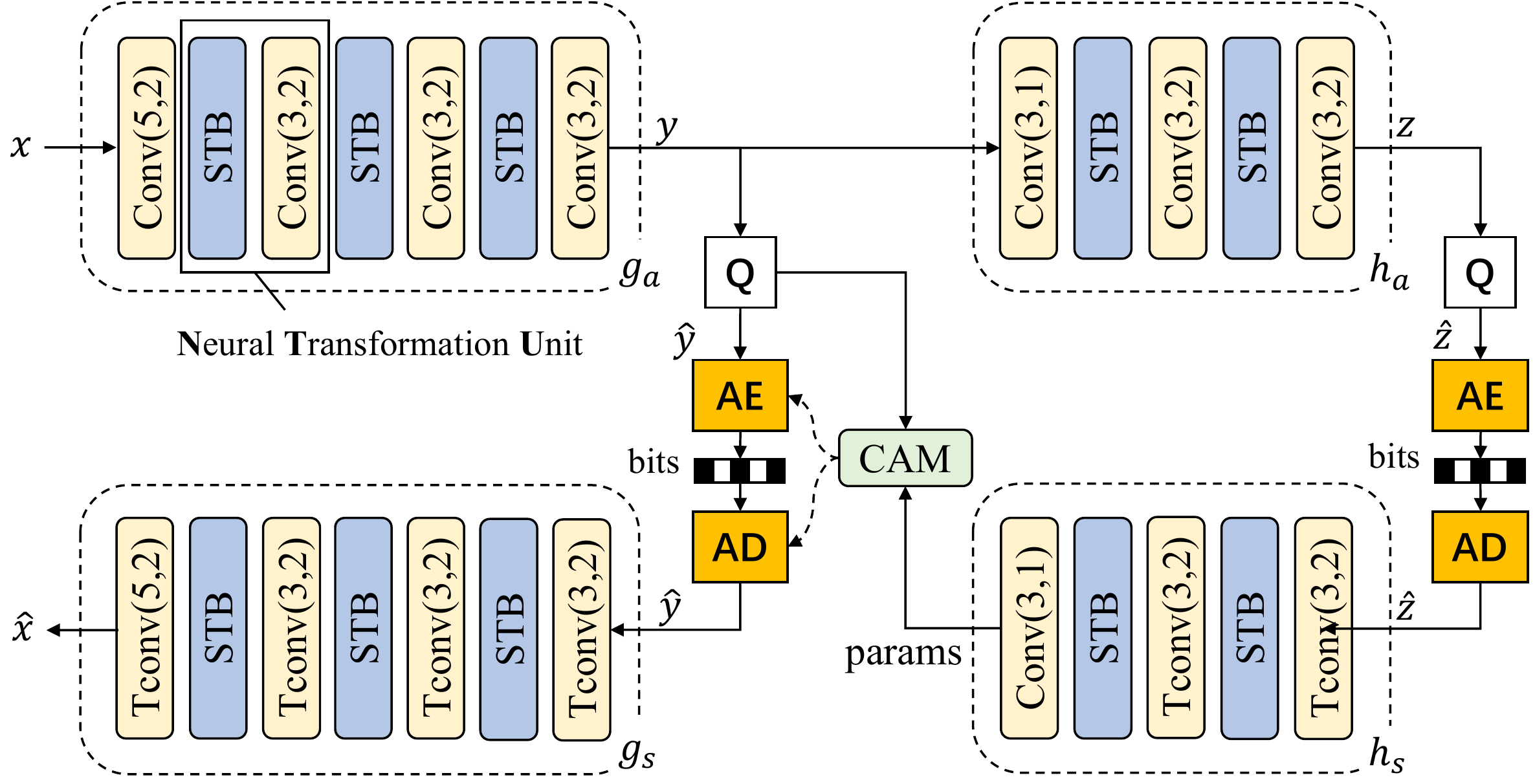}
\end{tabular}
\end{center}
\caption{{\bf TIC.} The proposed TIC stacks the Neural Transformation Unit (NTUs) in both main and hyper encoder-decoders of a VAE framework. Each NTU is comprised of a STB and a Conv layer for which we wish to capture and embed long-range and short-range information. Conv(k,s) and Tconv(k,s) are convolutional and transpose convolutional layers with kernel size $k\times k$  and stride size $s$. $k$ = 3 and $s$ = 2 are exemplified in this work. The causal attention module (CAM) is applied to aggregate information from autoregressive neighbors and hyper priors for context modeling which is different from existing masked CNN based information fusion mechanism.} \label{fig:network}
\end{figure}

\begin{figure}
\begin{center}
\begin{minipage}[b]{0.8\linewidth}
\begin{center}
\subfloat[][STB]{
\label{fig:stb}
\includegraphics[width=0.40\linewidth]{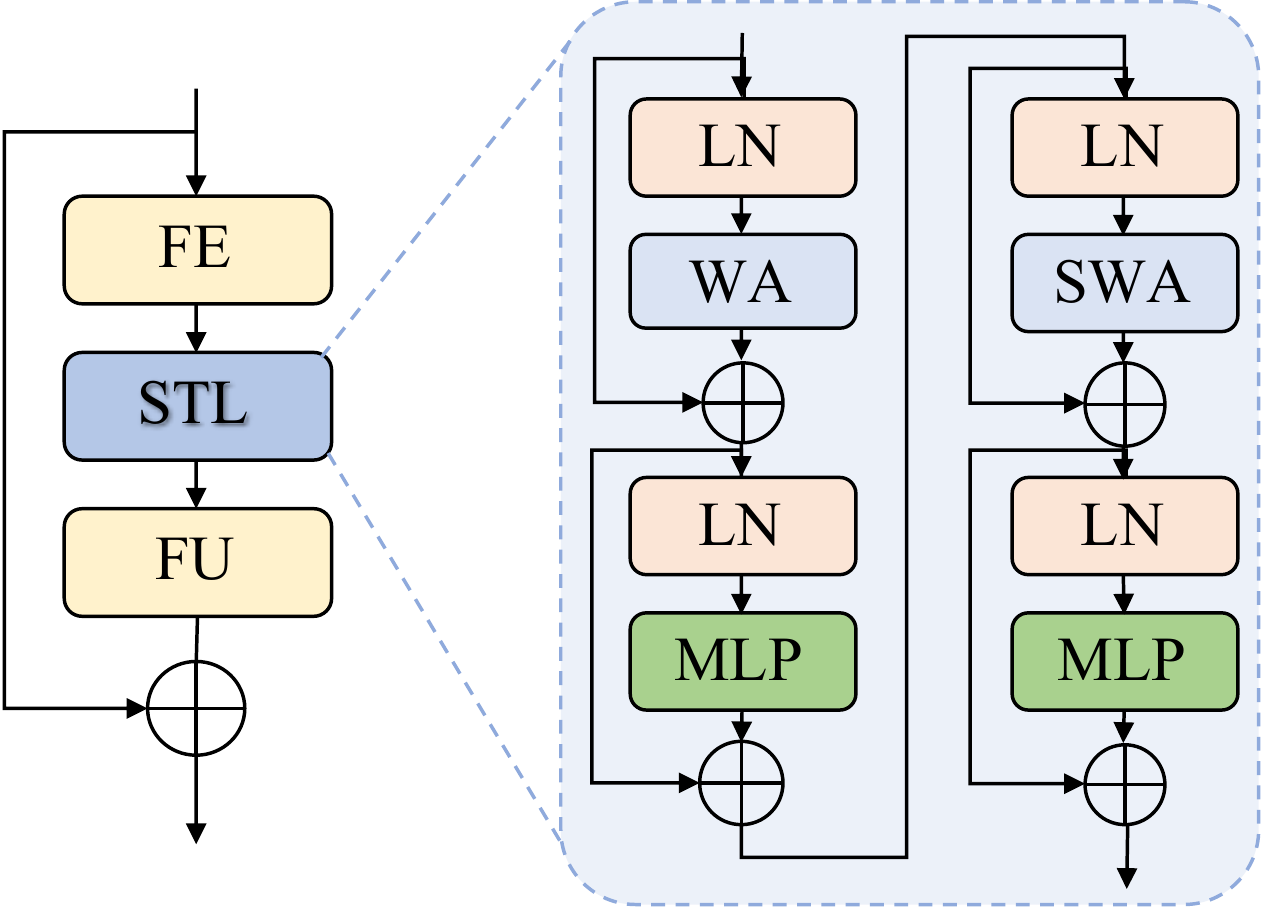}
}
\subfloat[][CAM]{
\label{fig:cam}
\includegraphics[width=0.55\linewidth]{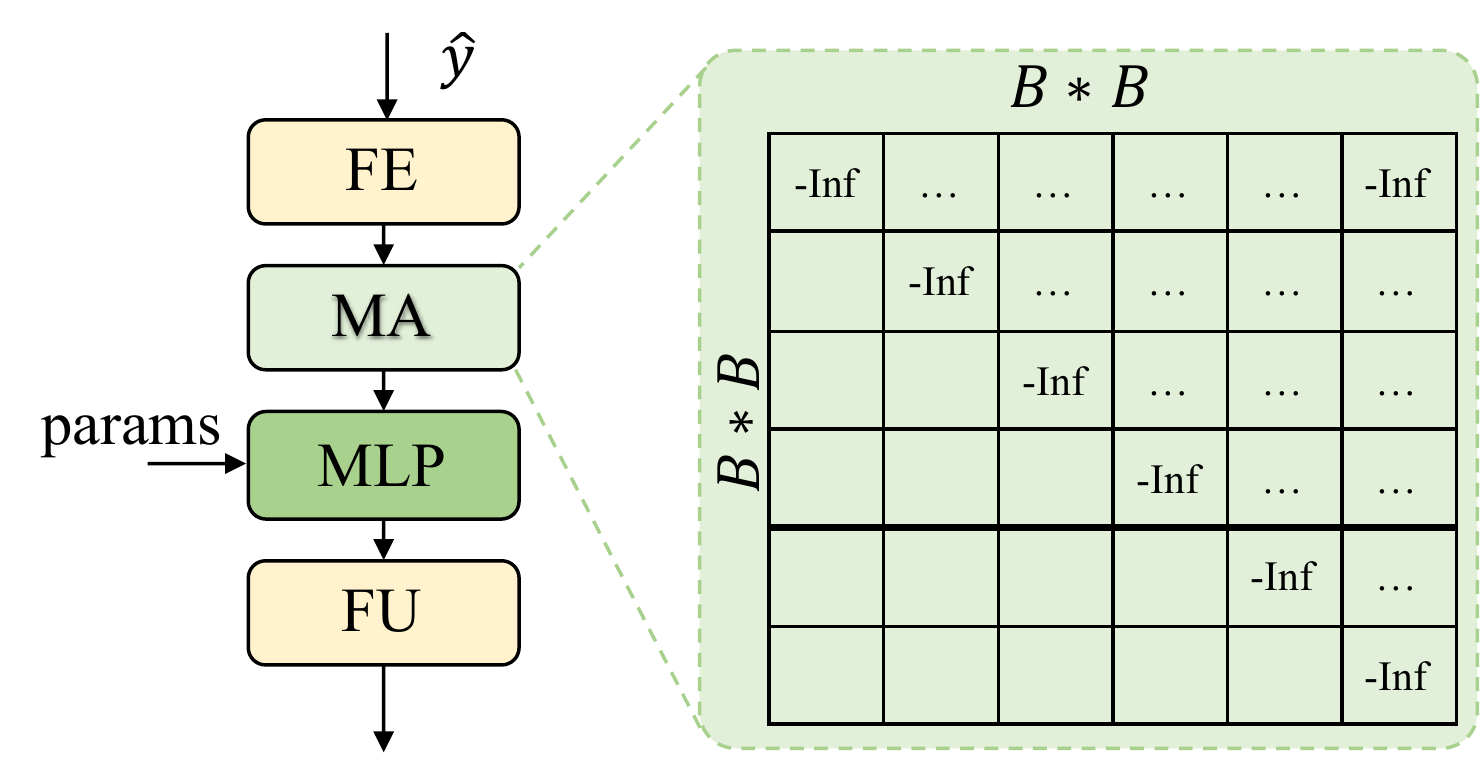}
}
\end{center}
\end{minipage} %\par
% \medskip
% \begin{minipage}[b]{.32\linewidth}
% \begin{center}
% \subfloat[][STL]{
% \label{fig:stl}
% \includegraphics[width=0.9\linewidth]{Figures/stl.pdf}
% }
% \end{center}
% \end{minipage}
% \begin{minipage}[b]{.32\linewidth}
% \begin{center}
% \subfloat[][MA]{
% \label{fig:ma}
% \includegraphics[width=0.75\linewidth]{Figures/mca.pdf}
% }
% \end{center}
% \end{minipage}
\caption{{\bf TIC Modules.} (a) Swin Transformer Block (STB); (b) Causual Attention Module (CAM) for context modeling.}
\label{fig:modules}
\end{center}
\end{figure}

Figure~\ref{fig:network} details the diagram of the proposed TIC. We follow the canonical VAE architecture~\cite{balle2018variational} to construct main and hyper encoder-decoder pairs. For the main encoder $g_a$, input image is first convoluted prior to being fed into succeeding three NTUs. The hyper encoder $h_a$ shares similar architecture but with two NTUs.
The main decoder $g_s$ and hyper decoder $h_s$ reverse the processing steps of $g_a$ and $h_a$ respectively.
{Each NTU applies a STB and a Conv layer to step-wisely analyze and embed respective long-range and short-range information.} 

As in STB,  the first feature embedding (FE) layer projects input features at a size of $ H \times W \times C$ to a dimension of $HW \times C$, the following Swin Transformer layer (STL)  which consists of the layer normalization (LN), window attention (WA), shifted window attention (SWA) and the MLP layer calculates the window-based self-attention, and finally a  feature unembedding (FU) layer remaps  attention weighted features back to original size of $ H \times W \times C$. Skip connection is used for better aggregation. Note that no patch division and fully-connected layers for tokenization in our work are  beneficial to early visual processing and stable training~\cite{gulati2020conformer,xiao2021early}. 

As in~\cite{chen2021end}, resolution scaling is integrated in the Conv layer with stride 2, by which we further aggregate local blocks or tokens in a spatial neighborhood. 
Currently, we simply use the linear GDN and Leaky ReLU for activation after the Conv in main and hyper coders. Other simple activation functions such as the ReLU can be used as well.

%\textbf{In the meantime, by utilizing the convolutional layers as the token pooling layers, tokens processed by the window-based STB blocks can be better fused spatially.(I could't figure out this sentence.)} 

Different from existing masked CNN based context modeling, we propose a  causal attention module (CAM) in Fig.~\ref{fig:cam} to select attentive neighbors from  autoregressive priors where the CAM unfolds the quantized features into $B\times B$ patches and calculate relations among these patches with masked attention (MA) to guarantee the causality. Experiments suggest  $B$ = 5 for well balancing the performance and complexity. The succeeding MLP layer fuse attention weighted autoregressive neighbors and hyper priors from $h_s$ for final context prediction.  

We later extend the TIC by placing more STBs at layers closer to the bottleneck, i.e., having 1 STB, 2 STBs, and 3 STBs respectively in three NTUs of the main encoder-decoder, and keeping the same 1 STB in each NTU in hyper coder, named as the TIC+.

\begin{figure}[t]
\begin{center}
\subfloat[][]{
\label{fig:rd_comparison}
\includegraphics[width=0.5\linewidth]{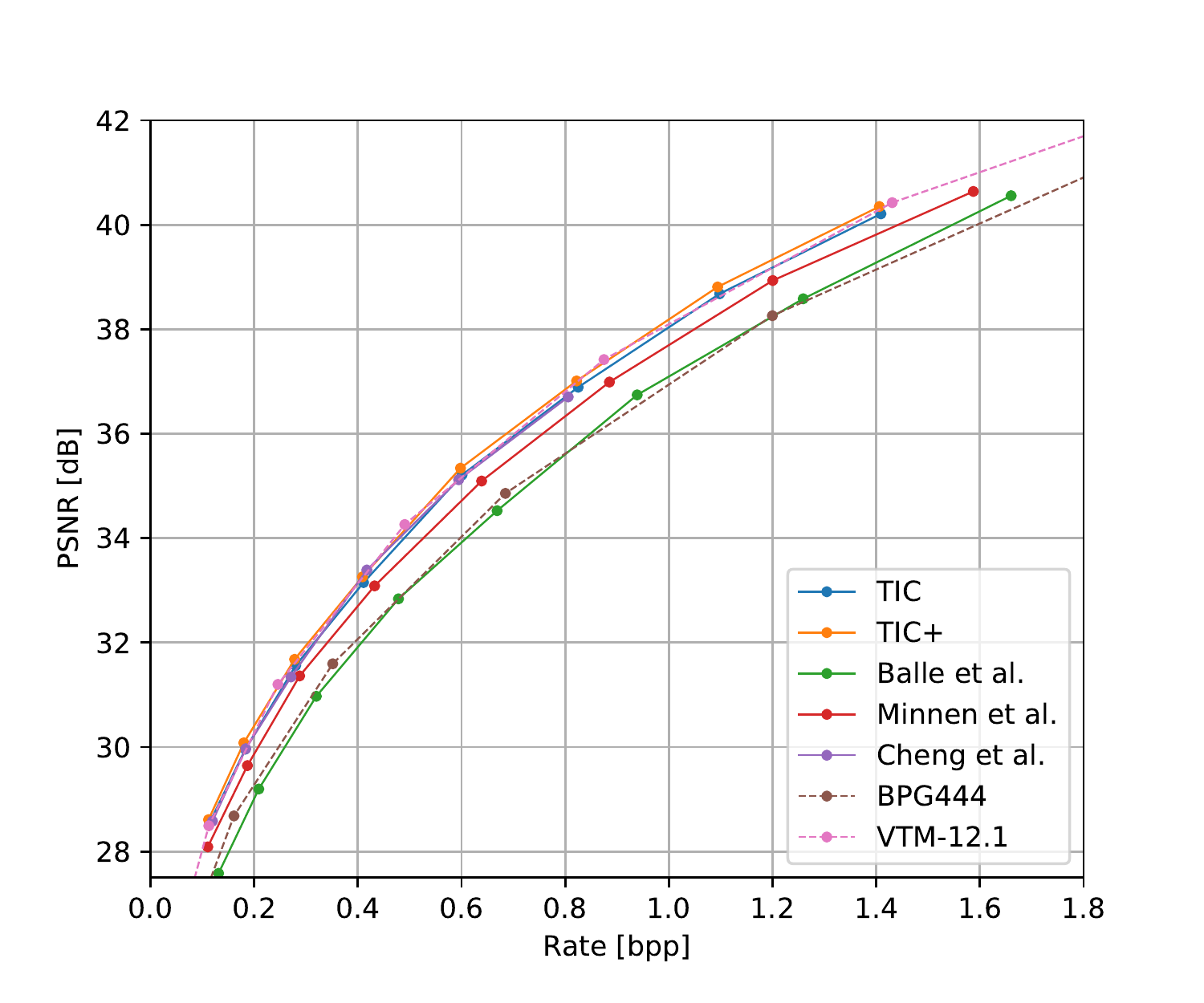}
}
\subfloat[][]{
\label{fig:param_comparison}
\includegraphics[width=0.5\linewidth]{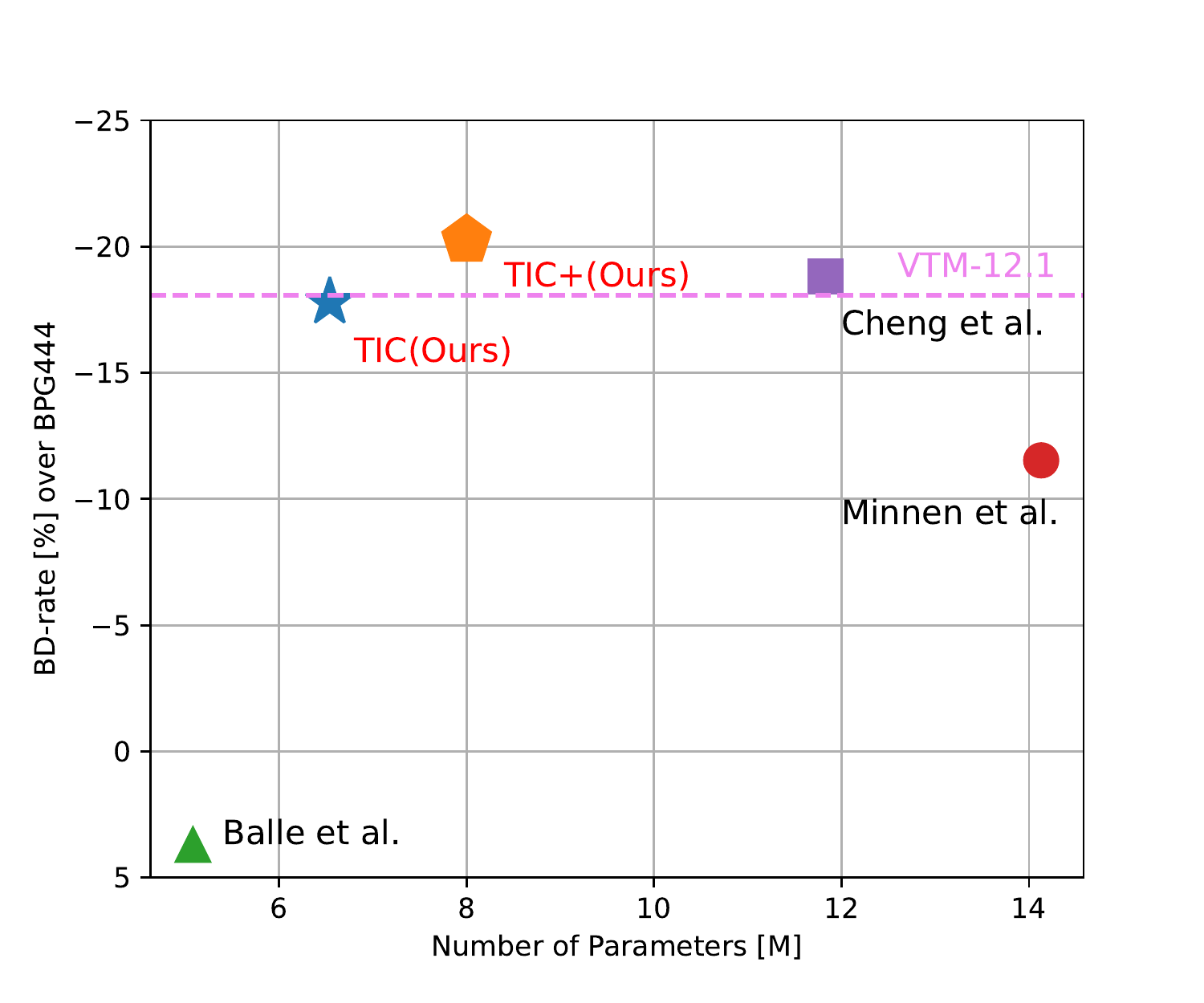}
}
\end{center}
\caption{{\bf Quantitative Evaluation.}
(a) R-D Performance of the TIC and TIC+ against the BPG444 (HEVC Intra), VTM 12.1 (VVC Intra) and other LICs with leading performance~\cite{balle2018variational,minnen2018joint,cheng2020learned}; (b) Performance gains (BD-Rate) against the HEVC Intra using BPG444 mode, and model parameters (in Mbytes).  Upper left is better. Note that performance is averaged using all test images in Kodark dataset.} \label{fig:comparison} 
\end{figure}

\begin{figure}[t]
\begin{center}
\begin{tabular}{c}
\epsfig{width=\linewidth,file=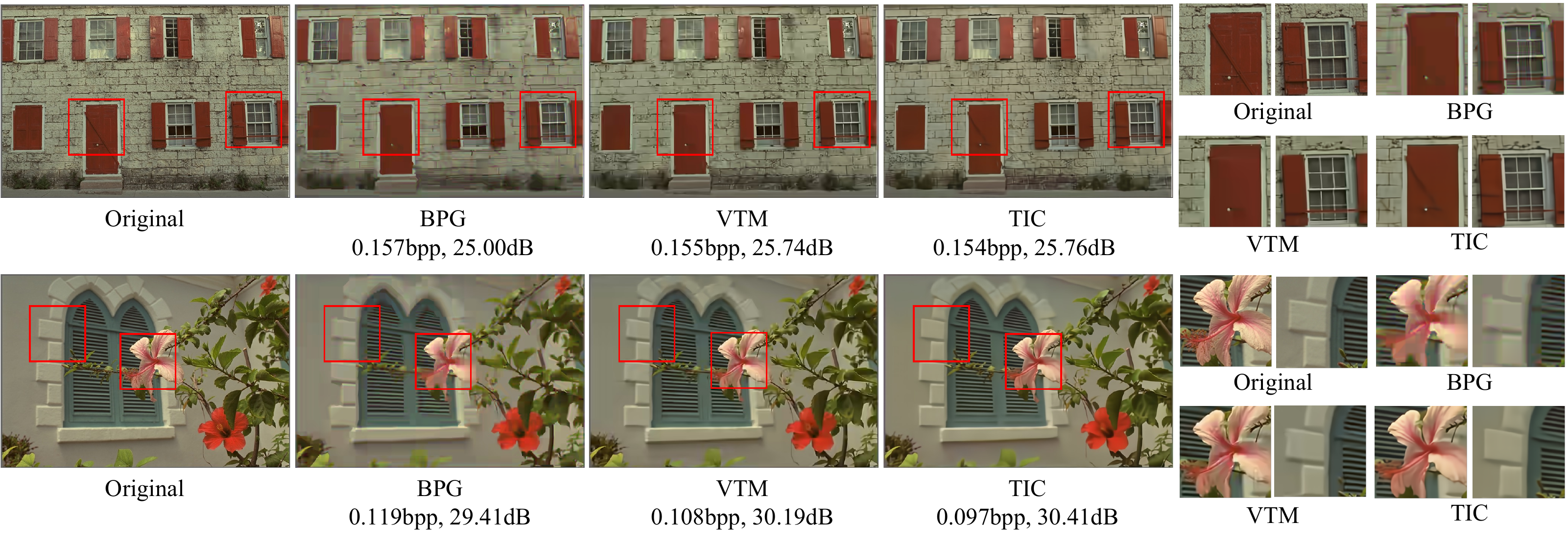}
\end{tabular}
\end{center}
\caption{{\bf Qualitative Visualization.}
Reconstructions and close-ups of the TIC, BPG (HEVC Intra) \& VTM (VVC Intra). Corresponding bpp and PSNR are marked.} \label{fig:visual_compare}
\end{figure}

\begin{figure}
\begin{center}
\begin{minipage}[b]{1\linewidth}
\begin{center}
\subfloat[][kodim04]{
\label{fig:vis_org_04}
\includegraphics[width=0.285\linewidth]{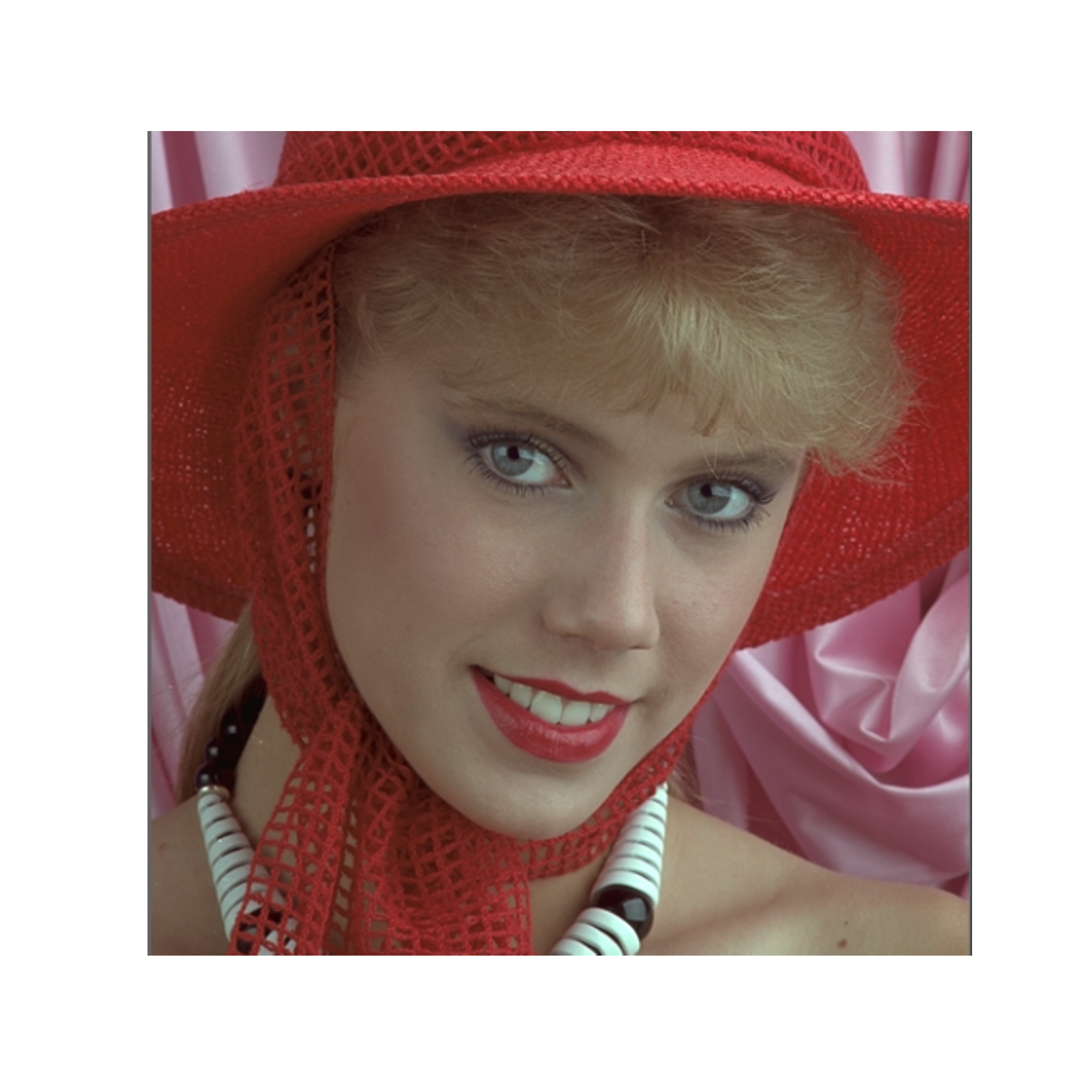}
}
\subfloat[][Cheng \etal]{
\label{fig:vis_cheng_04}
\includegraphics[width=0.33\linewidth]{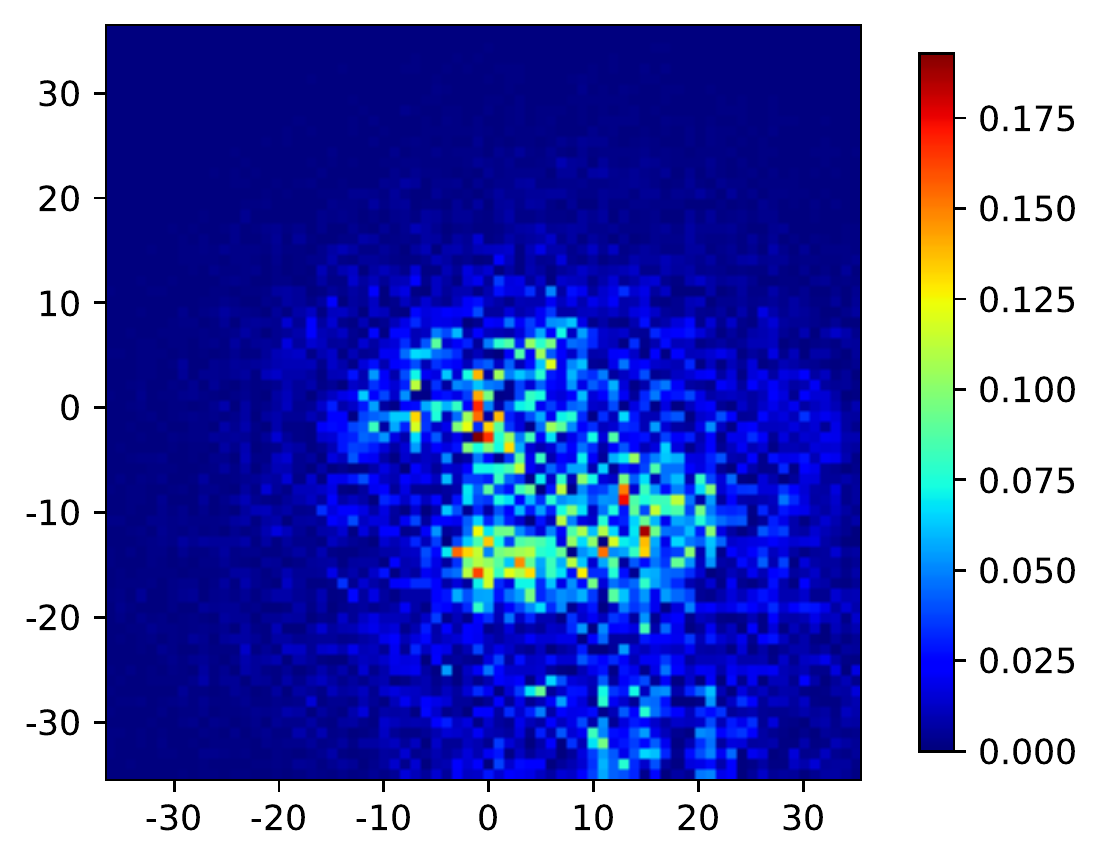}
}
\subfloat[][TIC]{
\label{fig:vis_tic_04}
\includegraphics[width=0.33\linewidth]{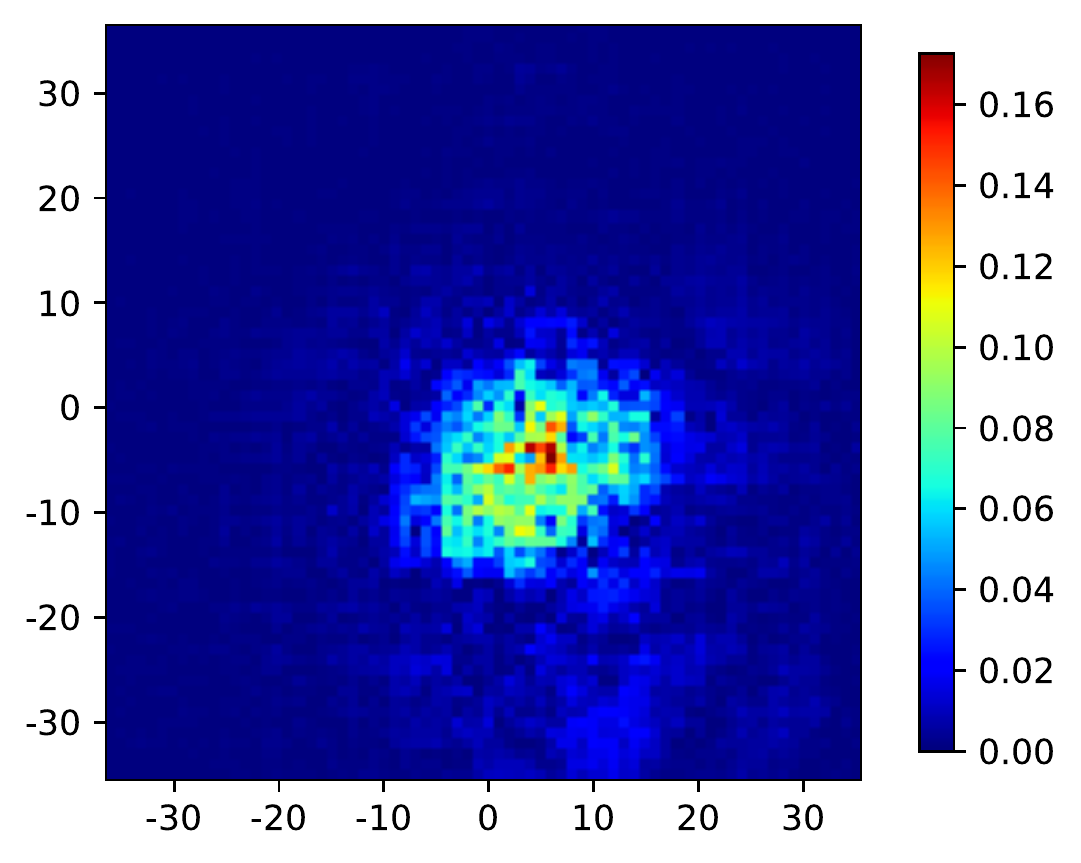}
}

\subfloat[][kodim24]{
\label{fig:vis_org_24}
\includegraphics[width=0.285\linewidth]{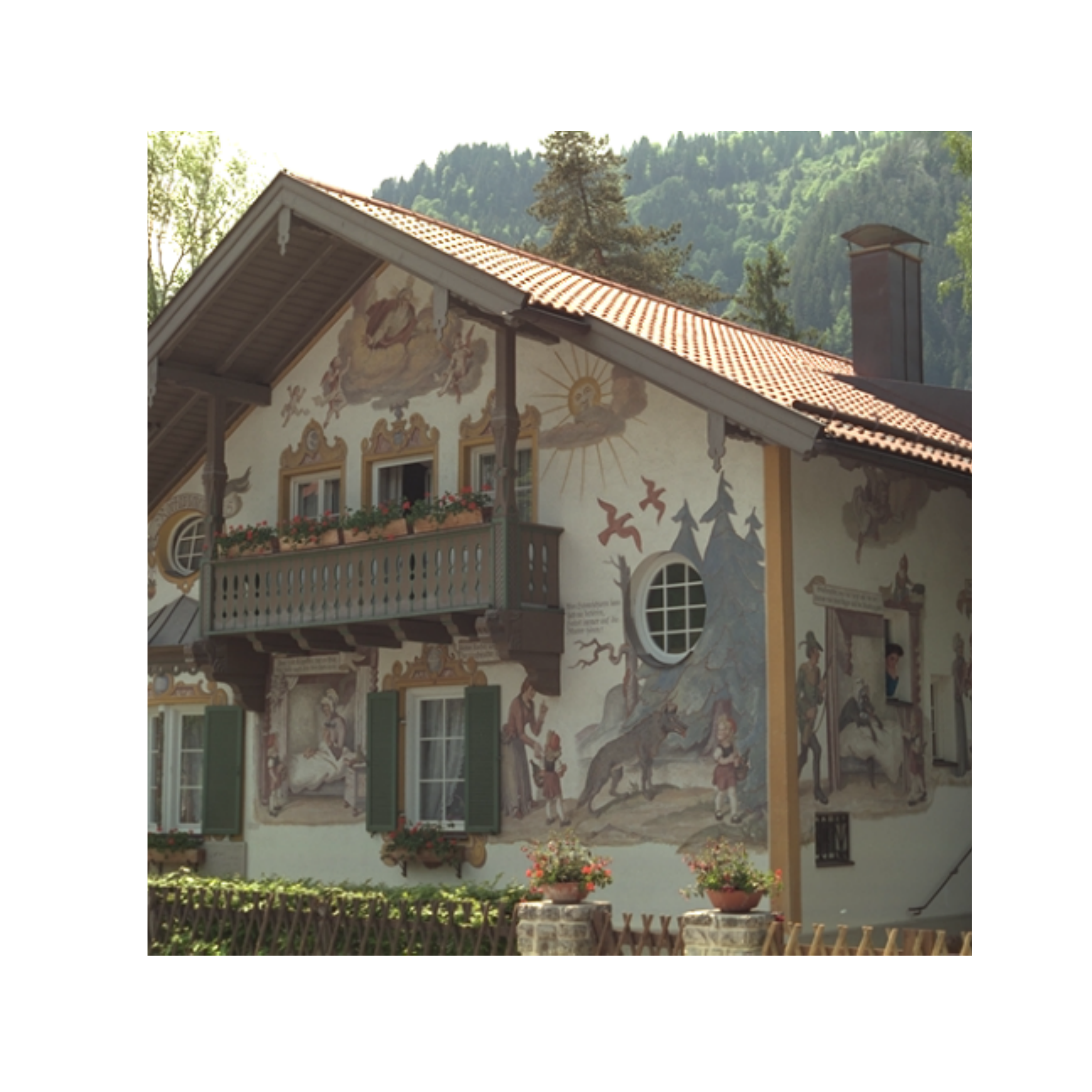}
}
\subfloat[][Cheng \etal]{
\label{fig:vis_cheng_24}
\includegraphics[width=0.32\linewidth]{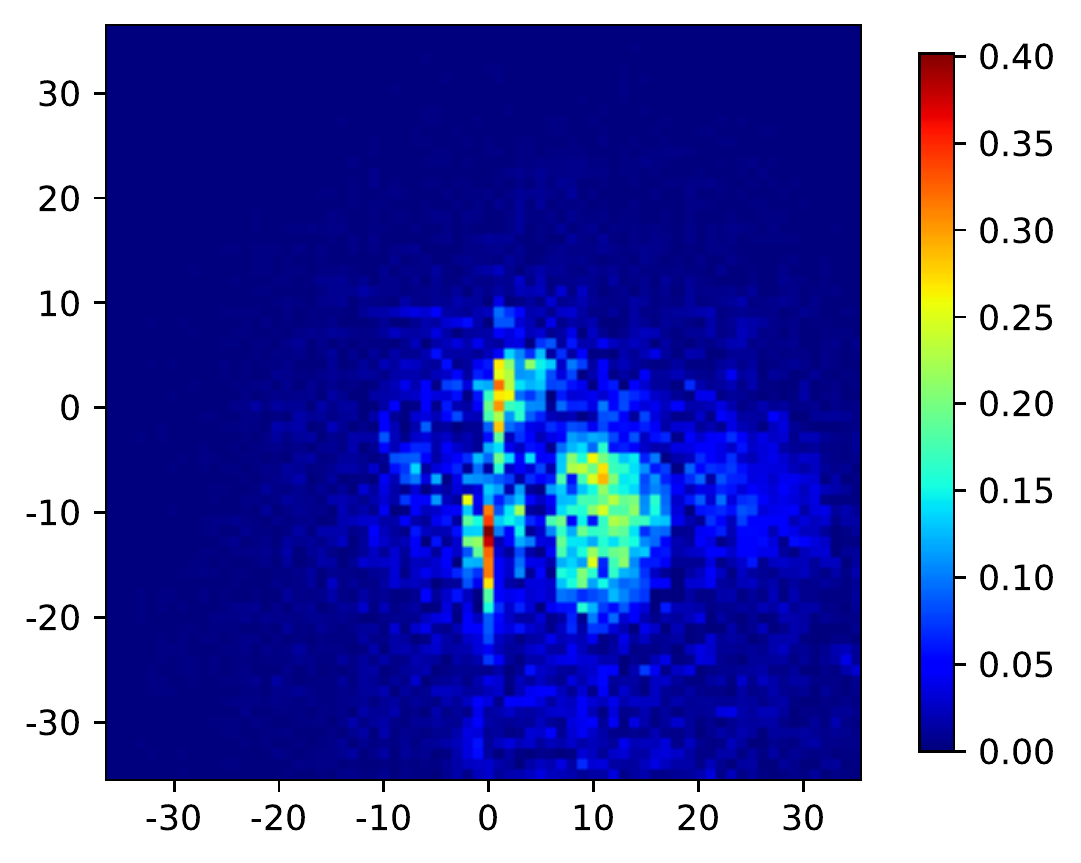}
}
\subfloat[][TIC]{
\label{fig:vis_tic_24}
\includegraphics[width=0.32\linewidth]{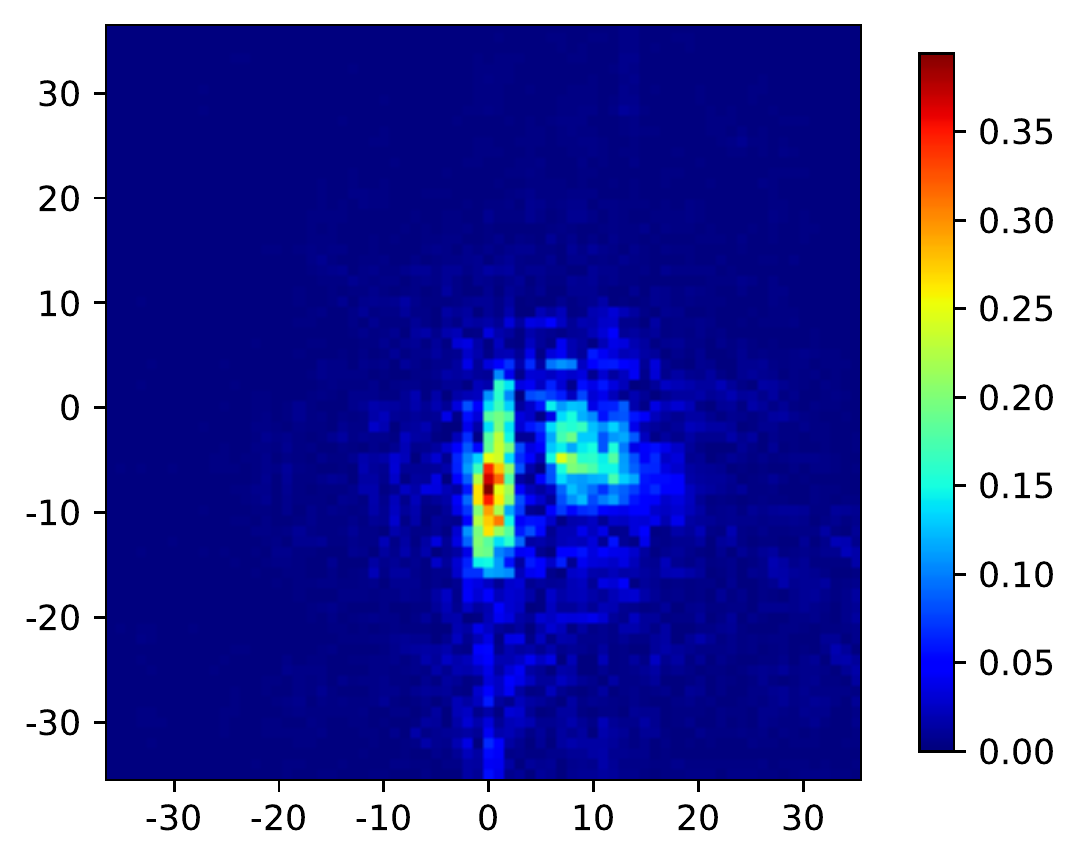}
}
\end{center}
\end{minipage} %\par
\caption{{\bf Compactness Visualization of Latent Features.}  Close-ups of the gradient maps averaged over all channels for different test images.}
\label{fig:visual}
\end{center}
\end{figure}

\section{Results}

{\bf Experimental Setup.} The Flicker 2W~\cite{liu2020unified} is used as the training dataset. We randomly crop images into fixed patches at a size of $256 \times 256 \times 3$. Adam is used as the optimizer and the batch size is set to 8 for training. All training threads run on a single Titan RTX GPU for 400 epochs with the learning rate of $10^{-4}$. We now train 8 models in total to match different bit rates (or quality levels) that can be adapted by selecting a  $\lambda$ is \{0.0018, 0.0035, 0.0067, 0.013, 0.025, 0.0483, 0.0932, 0.18\}. The MSE is used for distortion measurements. 

The proposed TIC is implemented on top of an open-source CompressAI PyTorch library~\cite{begaint2020compressai}, by which we can easily share our models and materials for reproducible research.  We evaluate our model on the Kodak dataset, having the peak signal-to-noise ratio (PSNR) to quantify the image quality and  the bits per pixel (bpp) to measure the bit rate.

%optimized with the mean squared error (MSE) metric under different quality levels, i.e., $\lambda$ is chosen from the set \{0.0018, 0.0035, 0.0067, 0.013, 0.025, 0.0483, 0.0932, 0.18\} in accordance with the recently well-developed 

In STB, the window sizes are set to 8$\times$8 and 4$\times$4 in respective main and hyper encoder-decoders and the numbers of heads are 4, 8, 16 for three STBs in main encoder-decoders and 16 for hyper STBs. We adopt 128 channels for the first 4 models corresponding to low bit rates scenarios, and 192 channels for the rest 4 models to cover high bit rates. 

%We also employ a version with more parameters called TIC+ by using 1, 2, 3, 1, and 1 STB layers according to the order for the encoder and inverse order for the decoder respectively.  

{\bf Quantitative and Qualitative Evaluations.}
Quantitative performance illustrated either using rate-distortion (R-D) curves in Figure~\ref{fig:rd_comparison} or using BD-Rate gains (over the HEVC Intra) in Figure~\ref{fig:param_comparison} reports the competitive efficiency of the proposed TIC to the state-of-the-art Cheng \etal~\cite{cheng2020learned}, and VVC Intra (VTM 12.1), but the TIC only requires a half of model parameters to Cheng \etal~\cite{cheng2020learned} which is more preferred for real-life application.

And, the TIC+ even surpasses the VVC Intra by 2.6\% BD-Rate improvement with a slight increase of model parameters. Even though, the TIC+ still consumes much less parameters than the  Cheng \etal~\cite{cheng2020learned}.

Figure~\ref{fig:visual_compare} visualize the reconstructions and closeups generated by the TIC, HEVC Intra (BPG), and VVC Intra (VTM), which agrees with the objective improvement of proposed TIC by subjective illustration with more sharp textures and less noise. 

{All of these studies have revealed the efficiency by combing the convolution and attention-based Transformer. By placing more STBs in TIC+, we observe more coding gains. This would be an interesting topic in future to study how many STBs are sufficient for achieving the optimal coding efficiency. }

{As aforementioned, the core issue of image compression is about the generation of compact representation of input image. We then use  the partial derivative $\frac{\partial g_a(m,n)}{\partial I(i,j)}$ to visualize  the compactness of input pixel $I(i,j)$ to the latent feature on position $(m,n)$ generated by the main encoder $g_a$. Figure~\ref{fig:visual} exemplifies the scenario when we take the center of $g_a$  to illustrate the contribution from all pixels of input image.} Apparently, our TIC offers much compact illustration than Cheng \etal's model that uses the CNN-based method, which further evidence that the combination of Transformer-based attention and convolution can better embed spatial information with more compact representation, and thus can lead to better coding efficiency as reported.

\section{Conclusion}
This paper reports a state-of-the-art image compression method. It combines the Swin Transformer and the convolutional layer as the basic unit to analyze and aggregate short-range and long-range information for more compact representation of input image. Experimental results reveal the leading performance when compared with existing learning-based approaches, and recently-emerged VVC Intra. Besides the encouraging coding efficiency, the proposed method consumes much less model parameters to the existing learning-based approaches, making the solution attractive to practical applications. An interesting topic for future study is to further improve the coding efficiency along the direction by stacking the convolutions and Transformers. More test results and models will be updated regularly at \url{https://njuvision.github.io/TIC}.

% \begin{figure}[t]
% \begin{center}
% \begin{tabular}{cc}
% \multicolumn{2}{c}{\epsfig{width=4in,file=Figures/image1}} \\
% \multicolumn{2}{c}{\small{(a)}} \\[1em]
% \epsfig{width=2in,file=Figures/image3.eps} &
% \epsfig{width=2in,file=Figures/image4.eps} \\
% {\small (b)} & {\small (c)}
% \end{tabular}
% \end{center}
% \caption{\label{fig:example}%
% An example figure.}
% \end{figure}

% \begin{table}[tp]
% \begin{center}
% \caption{\label{tab:example}%
% Average PSNR in dB for the ``Coastguard'' video sequence}
% {
% \renewcommand{\baselinestretch}{1}\footnotesize
% \begin{tabular}{|c|c|c|c|c|}
% \cline{2-5}
% \multicolumn{1}{c|}{~}&
% \multicolumn{1}{c|}{2D} &
% \multicolumn{1}{c|}{3D} &
% \multicolumn{2}{c|}{MC-BCS-SPL}\\
% \cline{4-5}
% \multicolumn{1}{c|}{$S_{\text{NK}}$} &
% BCS-SPL & BCS-SPL & $S_{\text{K}}=S_{\text{NK}}$ & $S_{\text{K}}=0.7$\\
% \hline
% 0.1 &22.69 &22.76 &23.06 &25.29 \\
% 0.2 &24.70 &24.76 &25.78 &27.94 \\
% 0.3 &26.37 &26.45 &28.29 &30.15 \\
% 0.4 &27.99 &27.95 &30.88 &32.30 \\
% 0.5 &29.60 &29.57 &33.58 &34.42 \\
% \hline
% \end{tabular}}
% \end{center}
% \end{table}

\Section{References}
\bibliographystyle{IEEEbib}
\bibliography{refs}

\end{document}